\documentclass[12pt,twoside]{article}
\usepackage[ascii]{inputenc}
\usepackage[T1]{fontenc}
\usepackage[english]{babel}
\usepackage{amsmath,amssymb,amsfonts,textcomp}
\usepackage{color}
\usepackage{calc}
\usepackage{longtable}
\usepackage{hyperref}

\input{tcilatex}

\begin{document}

\title{{Romania in a post{}-credit crunch world? A cautionary tale from
Australia and America}}
\author{{Carmen Costea (Academy of Economic Studies, Bucharest, Romania) and
Steve Keen (University of Western Sydney, Australia)}}

\begin{abstract}
We present data on debt accumulation in Australia and the United States, and
tentative data on Romania, to pose the question of whether Romania might
experience a credit crunch as a result of the US\ subprime financial crisis.
We develop a model of a credit crunch in a pure credit economy with
endogenous money creation, to show how changes in bank lending practices and
borrower repayment behaviour can bring about an economic decline.
\end{abstract}

\maketitle

{As this paper goes to press, America's Dow Jones stockmarket index is
officially in a \textquotedblleft correction\textquotedblright ---defined as
a fall of more than ten percent fall from a previous peak. More ominously,
American house prices are more than eight percent below their peak, and are
currently falling at a rate that exceeds one percent per month{---}an
unprecedented rate of decline.}

\FRAME{ftbphFU}{4.0413in}{2.437in}{0pt}{\Qcb{Dow Jones Industrial\ Average 
\protect\cite{DJIA} \&\ Case-Schiller Housing Price Index \protect\cite%
{CaseSchiller}}}{\Qlb{Figure01USA_AssetBubbles}}{01usaassetbubbles.wmf}{%
\special{language "Scientific Word";type "GRAPHIC";maintain-aspect-ratio
TRUE;display "USEDEF";valid_file "F";width 4.0413in;height 2.437in;depth
0pt;original-width 5.4691in;original-height 3.2811in;cropleft "0";croptop
"1";cropright "1";cropbottom "0";filename
'01USAAssetBubbles.wmf';file-properties "XNPEU";}}

{The proximate cause of this shakeout in American asset markets is the
so{}-called \textquotedblleft subprime lending crisis\textquotedblright .
Professional opinions differ on how long this crisis will last and how
serious its consequences will be, but the minimum expectation is that a
recession as colloquially defined (two quarters or more of negative growth)
will result.}

{In this paper we suggest that the consequences could be much more severe.
The basis of our pessimism is a closer look at the broader phenomenon of
which the subprime crisis is merely the latest installment: a multi{}-decade
trend for private debt to rise faster than income.}

{Research by Australia's central bank, the Reserve Bank of Australia
(Battellino 2007), indicates that over the three decades from 1977 to 2007,
private debt (the sum of business and household debt) has risen faster than
nominal GDP in 15 of the major OECD nations (the only significant exception
is France).}

{On this gauge, the US debt phenomenon no longer appears especially
remarkable. Though the USA's ratio of private debt to GDP more than doubled
over those three decades, this increase was at the low end of the
international scale, which ranged from a maximum of an elevenfold increase
for The Netherlands to a doubling for Germany (Japan is a special case,
which we discuss later). The median experience has been that of Australia,
whose private debt to GDP ratio has risen threefold in the last 30 years{---}%
and sixfold since the mid{}-1960s. The rise in private indebtedness is
therefore a global phenomenon{---}which we argue is unsustainable, and must
at some point reverse. When it does, the economic circumstances will be very
different to those that have applied for the past three decades. We will
pass from a debt{}-driven economy to one dominated by a \textquotedblleft
credit crunch\textquotedblright .}

\FRAME{ftbphFU}{4.0413in}{4.0413in}{0pt}{\Qcb{RBA\ research on private debt
to GDP across the OECD \protect\cite{Battellino2007}}}{\Qlb{%
Figure02Debt2GDP_OECD}}{02rbadebt2gdpdata.jpg}{\special{language "Scientific
Word";type "GRAPHIC";maintain-aspect-ratio TRUE;display "USEDEF";valid_file
"F";width 4.0413in;height 4.0413in;depth 0pt;original-width
4.6458in;original-height 4.6458in;cropleft "0";croptop "1";cropright
"1";cropbottom "0";filename '02RBADebt2GDPData.wmf';file-properties "XNPEU";}%
}

\section{From the mid{}-60s till 2008: the debt{}-driven economy}

{The growth in debt illustrated by the above charts has a profound impact
upon economic performance that is neglected by conventional economic
analysis, which ignores the role of monetary factors in economics.}

\FRAME{ftbphFU}{4.0413in}{3.2949in}{0pt}{\Qcb{USA data \protect\cite{FRB};
Australian data \protect\cite{RBA}}}{\Qlb{USA_Aus_Debt2GDP}}{%
03aususadebt2gdp.jpg}{\special{language "Scientific Word";type
"GRAPHIC";maintain-aspect-ratio TRUE;display "USEDEF";valid_file "F";width
4.0413in;height 3.2949in;depth 0pt;original-width 4.6458in;original-height
3.781in;cropleft "0";croptop "1";cropright "1";cropbottom "0";filename
'03AUSUSADebt2GDP.wmf';file-properties "XNPEU";}}

{Aggregate demand in an economy for everything from commodities to assets{---%
}technically, Gross National Expenditure{---}is the sum of income and the
change in debt. When debt levels are low compared to GDP, the contribution
from change in debt is negligible, and this monetary side of the economy can
comfortably be ignored. But when debt levels become large compared to GDP,
much of the effective demand{---}and the majority of the volatility in
economic performance{}-{}-is driven by changes in debt. Monetary factors are
no longer economically irrelevant. Australia}'{s recent economic history
clearly illustrates this shift from an economy where monetary factors can be
ignored, to one where they are dominant.}

{As noted above, Australia}'{s debt to GDP history is closer to the mean for
all OECD countries than that of the USA. Also, even though Australia}'{s
economy is currently performing well, while America}'{s is now feared to be
in recession after the collapse of the subprime market, Australia}'{s
economic history also provides a clearer example than America of a
speculation{}-dominated economy.}

{In the 1950s and early 1960s, when Australia}'{s private debt was less than
25\% of GDP, annual changes in debt contributed less than 5 percent to Gross
National Expenditure. That changed markedly in 1973, when the first and
smallest of three recent {"}super{}-bubbles{"} in debt occurred (see Figure
3). Briefly, the change in debt that year accounted for 10 percent of
nominal GNE (see Figure 4).}

{Simultaneously, the great 1970s inflationary surge began, and unemployment
in Australia exploded from its previous historic level of below two percent,
to over six percent. Most economists blamed the downturn on poor government
economic policy and inflation, but the real cause was the collapse in
speculative bubble that the growth in debt had financed. With the bubble
over, speculators shifted from willingly taking on debt, to trying
desperately to reduce it. The rate of growth of debt fell well below the
rate of inflation, and the change in debt went from boosting real aggregate
demand to subtracting from it. The Australian economy went into its first
post{}-WWII recession{---}caused by a decrease in the rate of growth of
private debt.}

\FRAME{ftbphFU}{4.0413in}{3.1514in}{0pt}{\Qcb{Data source: \protect\cite{RBA}%
}}{\Qlb{AusDebtDemandInflation}}{04debtchangedemandinflation.jpg}{\special%
{language "Scientific Word";type "GRAPHIC";maintain-aspect-ratio
TRUE;display "USEDEF";valid_file "F";width 4.0413in;height 3.1514in;depth
0pt;original-width 5.3956in;original-height 4.1978in;cropleft "0";croptop
"1";cropright "1";cropbottom "0";filename
'04DebtChangeDemandInflation.wmf';file-properties "XNPEU";}}

{The dependence on debt became even more extreme as debt rose from 44\% of
GDP in 1973 to 85\% in 1990, during the next debt super{}-bubble. At the
peak of that bubble, increases in private debt accounted for almost 14\% of
aggregate demand. This was a huge credit{}-driven boost to the economy while
it lasted, but when it went into reverse the change in debt turned negative,
and reductions in debt reduced nominal GNE by almost 1.5 percent. Real GNE
fell by substantially more, since inflation dropped along with the fall in
debt. The economy entered into its deepest post{}-WWII recession yet, with
unemployment exceeding 11 percent.}

{We face the dilemma that with high debt levels, economic performance
becomes dependent on the further accumulation of debt. When debt is small,
or changing by only small amounts, most variation in economic performance is
due to real productive factors. But when debt is much larger than output,
changes in debt contribute disproportionately to changes in apparent
economic performance. Ironically, superficially good economic performance{---%
}such as falling unemployment{}-{}-becomes dependent on ultimately
unsustainable further increases in debt.}

{This is apparent in the correlation between unemployment and change in debt
in the Australian data. In the 50s and 60s, when debt was under 25\% of GDP,
changes in debt made a comparatively small contribution to changes in
effective demand, and hence the correlation between changes in debt and
changes in unemployment was small (and positive).}

\FRAME{ftbphFU}{4.0413in}{3.2603in}{0pt}{\Qcb{Data source: \protect\cite{RBA}%
}}{\Qlb{AusChangeDebtUnemployment}}{05changedebtunemployment.jpg}{\special%
{language "Scientific Word";type "GRAPHIC";maintain-aspect-ratio
TRUE;display "USEDEF";valid_file "F";width 4.0413in;height 3.2603in;depth
0pt;original-width 6.1462in;original-height 4.9476in;cropleft "0";croptop
"1";cropright "1";cropbottom "0";filename
'05ChangeDebtUnemployment.wmf';file-properties "XNPEU";}}

{However, as debt rose relative to GDP, changes in debt made a much larger
contribution to changes in effective demand, and hence to changes in
unemployment. The correlation between changes in debt and changes in
unemployment consequently increased in magnitude as debt accumulated over
time. Now, that correlation has stabilized at more than {}-90
percent{}-{}-which means that a decline in the rate of growth of debt is
highly likely to be correlated with an increase in the level of unemployment
(see Figures 5 and 7).}

\FRAME{ftbphFU}{4.0413in}{3.2603in}{0pt}{\Qcb{Data source: \protect\cite{FRB}%
.}}{\Qlb{USA_ChangeDebtUnemployment}}{06changedebtunemploymentusa.jpg}{%
\special{language "Scientific Word";type "GRAPHIC";maintain-aspect-ratio
TRUE;display "USEDEF";valid_file "F";width 4.0413in;height 3.2603in;depth
0pt;original-width 6.1462in;original-height 4.9476in;cropleft "0";croptop
"1";cropright "1";cropbottom "0";filename
'06ChangeDebtUnemploymentUSA.wmf';file-properties "XNPEU";}}

{A similar observation applies to the American economy (see Figures 6 and
7). The correlation of changes in private debt with unemployment is more
volatile than in the Australian case{---}in part because of the
comparatively massive contributions from changes in government and financial
sector debt to American demand. But clearly the Australian and American
economies, in concert with 13 other major OECD nations, have become {"}%
addicted to debt{"}.}

\FRAME{ftbphFU}{4.0413in}{3.4688in}{0pt}{\Qcb{Data source: \protect\cite{RBA}%
}}{\Qlb{AusCorrelationUnempDebtChange}}{07corrunempchangedebt.jpg}{\special%
{language "Scientific Word";type "GRAPHIC";maintain-aspect-ratio
TRUE;display "USEDEF";valid_file "F";width 4.0413in;height 3.4688in;depth
0pt;original-width 5.0211in;original-height 4.3024in;cropleft "0";croptop
"1";cropright "1";cropbottom "0";filename
'07CorrUnempChangeDebt.wmf';file-properties "XNPEU";}}

{This is the most conventionally economic danger facing the world economy as
the US subprime crisis spreads. As households go from willingly taking on
more debt to trying to reduce their indebtedness, the change in debt will go
from boosting aggregate demand to subtracting from it. \ The 1990s collapse
in debt levels caused unemployment in Australia to rise from six to eleven
percent{---}yet in 1990, the change in debt was responsible for {"}only{"}
14 percent of GNE. Today, it accounts for over 18 percent of GNE.}

{Similarly in the USA, when the debt{}-driven component of nominal GNE
dropped from 10 to 2 percent across the 1990s recession, unemployment rose
from 5 to almost 8 percent. Today, increasing debt accounts for 14 percent
of US GNE{---}and the subprime crisis clearly marks the end of this latest
and biggest debt bubble. The macroeconomic impact of the switch from
expanding to contracting debt levels is likely to result in the deepest
recession in the USA}'{s post{}-WWII economic history.}

\FRAME{ftbphFU}{4.0413in}{3.0952in}{0pt}{\Qcb{Data sources: \protect\cite%
{OECDData}, \protect\cite{FRB}, \protect\cite{RBA}}}{\Qlb{RomaniaDebt}}{%
08romaniadebt.jpg}{\special{language "Scientific Word";type
"GRAPHIC";maintain-aspect-ratio TRUE;display "USEDEF";valid_file "F";width
4.0413in;height 3.0952in;depth 0pt;original-width 4.9476in;original-height
3.781in;cropleft "0";croptop "1";cropright "1";cropbottom "0";filename
'08RomaniaDebt.wmf';file-properties "XNPEU";}}

{Since so much of the OECD is in a similar debt{}-driven state to America,
the turnaround in debt that is occurring there is likely to be replicated
across much of the OECD. It is therefore highly likely that at least an
OECD{}-wide recession will occur, if not a global recession. If so, this
will be the first time in Romania}'{s post{}-Revolution economic history
that external economic conditions have been contractionary rather than
expansionary. What are the implications of this for Romania itself?}

{Obviously this depends to some extent on the level of indebtedness in
Romania. Here, definitive and timely data is difficult to come by. The best
we could locate was the OECDSTAT database
(http://stats.oecd.org/wbos/\allowbreak
Default.aspx?usercontext=sourceoecd), where annual data to 2006 implies that
Romania}'{s aggregate debt to GDP ratio is even worse than that for the USA,
both in terms of magnitude and its rate of increase. This suggests that
Romania could face a debt{}-driven downturn in aggregate demand when the
rate of change of debt falls, in addition to a diminution in export demand.}

{Long term Australian economic data implies that this will be no ordinary
recession. Australia}'{s Reserve Bank assembled a long term data series on
the debt to GDP ratio, which shows that Australia}'{s economic performance
since 1965 has been driven by a debt bubble which is the third and by far
the biggest in its economic history.}

\FRAME{ftbphFU}{4.0413in}{3.0917in}{0pt}{\Qcb{Data sources: \protect\cite%
{Battellino2007}, \protect\cite{FisherKent1999}}}{\Qlb{AusLongTermDebt}}{%
09auslongtermdebtratio.jpg}{\special{language "Scientific Word";type
"GRAPHIC";maintain-aspect-ratio TRUE;display "USEDEF";valid_file "F";width
4.0413in;height 3.0917in;depth 0pt;original-width 5.2814in;original-height
4.0309in;cropleft "0";croptop "1";cropright "1";cropbottom "0";filename
'09AusLongTermDebtRatio.wmf';file-properties "XNPEU";}}

{It is likely that the same applies to the USA: that today}'{s private debt
levels are the largest in the history of capitalism, and that what we are
experiencing now is a repeat of the processes that gave rise to earlier
financial panics.}

{As Mark Twain famously remarked, {"}history doesn}'{t repeat, but it sure
does rhyme{"}, and Australia}'{s economic history gives strong reason to
expect something far more severe than a mere recession. The bursting of
Australia}'{s two previous historic debt bubbles{}-{---}in 1892 and
1931{}-{}-ushered in not merely recessions, but Depressions (long term data
on real GDP growth, money supply change and inflation implies that the USA
also experienced a Depression in the 1890s{}-{}-see Keen 2008A). Yet the
debt levels then were substantially less than today}'{s. It appears that the
existence of Central Banks that, as in the case of the US}'{s Federal
Reserve, respond to financial crises by trying to {"}save the private sector
from itself{"}, may actually have contributed to a {"}moral hazard{"}
dilemma that has allowed debt levels to exceed previous bounds.}

{If history is any guide, then without Central Bank activism{---}{}-such as
the rescue of Long Term Capital Management, and the aggressive cutting of
reserve rates during the 1990s recession{---}it is likely that the 1990s
bubble would have marked the peak of private debt accumulation (at 125
percent of GDP for the USA, and 85 percent for Australia). That in itself
would have been bad enough, given the historical record{---}given the
aftermaths of the previous two bubbles. But with today}'{s debt levels, we
truly are in unprecedented territory.}

{We are not saying that a global Depression is inevitable, because there are
other aspect of the modern economic and social system that differ
substantially to that of 1930 and 1890. Bank collapses and the destruction
of depositors savings, which were a feature of the 1890s in Australia (and
the 1930s in America), will not recur; and government social security
payments during a downturn will provide \ households with cash flows that
can be used to service debt, something that did not happen in 1930.}

{But there will inevitably be an extended period of reduced vitality to
aggregate demand, as income is channeled to pay debt levels down from today}'%
{s unprecedented levels to something closer to the 20{}-60 percent of GDP
level that appears sustainable in the long run. In Australia and America}'{s
cases, such a reduction in private debt would require more than an entire
year}'{s GDP to be directed simply at debt reduction. Since that cannot be
done in one hit, this implies a long period where demand will grow more
slowly than capacity to produce output rises.}

\section{The Ponzi Credit Dynamic}

{The Australian long term data indicates that there is something systemic in
Western economies that leads to periodic debt explosions, and subsequent
serious debt{}-driven downturns. The most cogent theory to explain this
phenomenon is the {"}Financial Instability Hypothesis{"}, which was
developed by the American economist Hyman Minsky in the late 1950s and early
1960s (see Keen 1995, 2008B). A key aspect of Minsky}'{s model was the
existence of {"}Ponzi financing{"}, in which individual speculators \ use
borrowed money to buy assets, and then attempt to profit by selling them to
other speculators for a higher price.\footnote{%
With very little irony, subprime lending can be summarized as a scheme to
make money by lending money to people who couldn{\textquotesingle}t afford
to repay it. It is thus a classic \textquotedblleft Ponzi
Scheme\textquotedblright , and the wonder is not that it collapsed, but that
anyone could take the Scheme seriously when it was first mooted.}}

{This behavior is dependent on asset prices rising faster than commodity
prices, and requires that debt rise faster still{---}and this has clearly
been the case in most OECD economies in the last 3 decades. This in turn
requires a financial system that willingly generates debt, up to a point at
which the debt burden causes a crisis that suddenly stems the flow of credit{%
---}a {"}credit crunch{"}.}

{Conventional models of money creation{}-{}-which argue that the banking
system does no more than amplify the monetary creation processes of the
Central Bank{---}cannot explain this process, and in any case fail to
account for the empirical data which shows that credit money creation
actually precedes the creation of Central Bank money by up to one year
(Kydland and Prescott 1990). A far more cogent explanation is given by the
Circuitist School model of endogenous money creation (Graziani 1989, 2004).
Recently one of us (Keen 2007) has developed a mathematical model of this
process in a pure credit economy.}

{The model explains how money is endogenously created, why banks are
motivated to extend credit indefinitely, and what happens to aggregate
demand when a credit crunch occurs. It is easily developed from a {"}%
double{}-entry book{}-keeping{"} table of the flows between accounts that
are initiated by a loan from a bank to a firm to finance production.}

{The model is developed in more detail in Keen 2008C (see also Chapman \&
Keen 2006). Here we will stick with a simpler presentation for the sake of
exposition. A bank loan to a firm creates two accounts: a record of debt 
\textit{F}${}_{\text{L}}$, and a deposit account for the firm \textit{F}${}_{%
\text{D}}$. An initial loan thus instantly creates a matching deposit, and
also sets up interest payment obligations between the bank and the firm: the
firm is obliged to pay interest on the outstanding debt, while the bank is
obliged to pay interest on the current level of the firm}'{s deposit account
(the rates differ of course, with the rate of interest on loans \textit{r}$%
{}_{\text{L}}$ exceeding that on deposits \textit{r}${}_{\text{D}}$.). The
funds flow between the firm}'{s and the bank}'{s deposit accounts, as shown
in the first row of Table One.}

{Table One: A model of endogenous money creation }

\begin{tabular}{|p{1.25cm}|p{1cm}|p{1cm}|p{1cm}|p{1.5cm}|p{1.5cm}|p{1.5cm}|p{1cm}|}
\hline
& \multicolumn{3}{|p{1cm}|}{Assets} & \multicolumn{4}{|c|}{Liabilities} \\ 
\hline
Account Type & Loans & Re\-s\-erves & Sum & \multicolumn{3}{|p{1.5cm}|}{
Deposits} & Sum \\ \hline
Name & $F_{L}$ & $B_{R}$ & $\dsum $ & $F_{D}$ & $B_{D}$ & $W_{D}$ & $\dsum $
\\ \hline
Interest &  &  &  & $r_{D}\cdot F_{D}\allowbreak -r_{L}\cdot F_{L}$ & $%
r_{L}\cdot F_{L}\allowbreak -r_{D}\cdot F_{D}$ &  & $0$ \\ \hline
Wages &  &  &  & $-\left( 1-s\right) \allowbreak \cdot P\cdot F_{D}$ &  & $%
\left( 1-s\right) \allowbreak \cdot P\cdot F_{D}$ & $0$ \\ \hline
Interest &  &  &  &  & $-r_{D}\cdot W_{D}$ & $r_{D}\cdot W_{D}$ & $0$ \\ 
\hline
Cons\-umption &  &  &  & $\beta \cdot B_{D}+\omega \cdot W_{D}$ & $-\beta
\cdot B_{D}$ & $-\omega \cdot W_{D}$ & $0$ \\ \hline
New \allowbreak Loans & $n_{M}\cdot F_{D}$ &  & $n_{M}\cdot F_{D}$ & $%
n_{M}\cdot F_{D}$ &  &  & $n_{M}\cdot F_{D}$ \\ \hline
Repay\-ment & $-R_{L}\cdot F_{D}$ & $R_{L}\cdot F_{D}$ & $0$ & $-R_{L}\cdot
F_{D}$ &  &  & $-R_{L}\cdot F_{D}$ \\ \hline
Re\-lending & $L_{R}\cdot B_{R}$ & $-L_{R}\cdot B_{R}$ & $0$ & $L_{R}\cdot
B_{R}$ &  &  & $L_{R}\cdot B_{R}$%
\end{tabular}%
\newline
\newline

{Once the firm has money it its deposit account, it can hire workers to
produce output. The flow of wages from firms to workers goes into the
workers deposit account \textit{W}${}_{\text{D}}$; this is the second row in
the table.}

{Workers therefore have positive bank balances, and they too receive
interest on these balances from the bank; this is the third row in the table.%
}

{Workers and bankers then buy commodities from the firm, resulting in
expenditure flows from the \textit{B}${}_{\text{D}}$ and \textit{W}${}_{%
\text{D}}$ accounts to the firm}'{s deposit account \textit{F}${}_{\text{D}}$%
: this is the fourth row in the table. The arguments \textrm{${\beta }$} and 
\textrm{${\omega }$} represent the rates of flow per annum out of each
account, relative to the balances at any time.}

{The next row in the table explains how money is endogenously created. In
the real world, firms negotiate lines of credit with banks, which enable
them to undertake expenditures{}-{}-and thus effectively create money in the
accounts of other firms{}-{}-which the banks record as a matching increase
in their outstanding debt levels. In the aggregate, this results in a
simultaneous increase in Firms deposits and firms recorded debt levels.
Unlike the previous rows in the table, this results in a net increase in
both bank liabilities{---}the sum of outstanding bank deposits{---}and bank
assets. The argument \textit{n}${}_{\text{M}}$ represents the rate at which
the money supply expands each year.}

{The next two rows record loan repayment flows from firms to banks, and the
flow of bank reserves from banks to firms{}-{}-effectively,
the\textquotedblleft recycling\textquotedblright\ of loans that have
previously been repaid. Both these transaction flows involve transfers from
the liability to the asset side of the bank}'{s ledger, but on the asset
side itself they simply result in the form of assets changing: from loans
(which generate an income flow to the bank) to reserves (which, being
inactive, do not generate an income flow to the bank). The arguments \textit{%
R}${}_{\text{L}}$ and \textit{L}${}_{\text{R}}$ represent the annual rate of
loan repayment and the annual rate of reserve recycling per annum
respectively.}

{We can now derive a dynamic model of endogenous money creation simply by
adding up the columns \ in the above table. Each column represents the flows
into and out of a given account. We thus have the following set of coupled
differential equations describing the basic dynamics of money creation in a
pure credit economy:}

\begin{eqnarray*}
\frac{d}{dt}F_{L} &=&n_{M}\cdot F_{D}-R_{L}\cdot F_{D}+L_{R}\cdot B_{R} \\
\frac{d}{dt}B_{R} &=&R_{L}\cdot F_{D}-L_{R}\cdot B_{R} \\
\frac{d}{dt}F_{D} &=&r_{D}\cdot F_{D}-r_{L}\cdot F_{L}-w\cdot F_{D}+\beta
\cdot B_{D}+\omega \cdot W_{D}+n_{M}\cdot F_{D}\allowbreak -R_{L}\cdot
F_{D}+L_{R}\cdot B_{R} \\
\frac{d}{dt}B_{D} &=&r_{L}\cdot F_{L}-r_{D}\cdot F_{D}-r_{D}\cdot
W_{D}-\beta \cdot B_{D} \\
\frac{d}{dt}W_{D} &=&w\cdot F_{D}+r_{D}\cdot W_{D}-\omega \cdot W_{D}
\end{eqnarray*}

{The model can now be used to explain why banks are predisposed to generate
as much credit as borrowers are willing to absorb up until a debt crisis
occurs, and also to show what happens when the system shifts from expanding
credit to a credit crunch.}

{On the first issue, bank income rises if the rate of money creation \textit{%
n}${}_{\text{M}}$ and the rate of loan recycling \textit{L}${}_{\text{R}}$
rise, while it falls if the rate of loan repayment \textit{R}${}_{\text{L}}$
rises. Banks therefore have a vested interest in increasing the rate of
money creation, increasing the rate of recicrulation of their reserves, and
discouraging borrowers from repaying loans.}

{Of course, as current economic conditions are now reminding us, this state
of affairs does not persist when loan defaults grow, and banks become
concerned that lending more money will lead not to more profits, but to
capital losses. The shift in sentiment we can now see in the USA, from
profligate lending to a credit crunch, involves a reversal in the above
three key parameters: banks reduce the rate at which they create new money,
the recirculation of existing reserves slows, and borrowers try to reduce
their indebtedness.}

{Figures 10 and 11 show this process with a doubling of the rate at which
borrowers attempt to repay loans, and a halving of both the rate of
recirculation of existing loans and of the rate of creation of new money.%
\footnote{%
The initial loan value is \$100, and parameter values before the credit
crunch are: \textit{r}${}_{\text{L}}$=5\%; \textit{r}${}_{\text{D}}$=3\%; 
\textrm{${\beta }$}=1; \textrm{${\omega }$}=26; \textit{n}${}_{\text{M}}$%
=10\% \textit{R}${}_{\text{L}}$=2; and \textit{L}${}_{\text{R}}$=2.} The
effect is a precipitous drop in money in circulation{}-{---}which
necessarily reduces the rate of economic activity.}

\FRAME{ftbphF}{4.0413in}{3.0882in}{0pt}{}{\Qlb{CreditCrunchDeposits}}{%
10depositscrunchmodel.jpg}{\special{language "Scientific Word";type
"GRAPHIC";maintain-aspect-ratio TRUE;display "USEDEF";valid_file "F";width
4.0413in;height 3.0882in;depth 0pt;original-width 4.6043in;original-height
3.5103in;cropleft "0";croptop "1";cropright "1";cropbottom "0";filename
'10DepositsCrunchModel.wmf';file-properties "XNPEU";}}

{Bank assets nonetheless continue to grow (though extending the model to
include bankruptcy would change this detail), but there is a dramatic shift
from active assets (loans) to inactive ones (reserves).}

{The system does stabilize and return to growth after a relatively short
period, but this is at a dramatically lower level of both active money and
economic activity.}

{Clearly, a process like this is currently underway in the USA, and given
the interconnectedness of the world financial system and the extent to which
the rest of the OECD is also debt{}-encumbered, it is only a matter of time
before the same process manifests itself worldwide. There will be
attenuating effects from countries which are net global creditors{---}such
as China and Japan{---}or which have not indulged in the orgy of Ponzi
financing (such as France), but these are unlikely to be sufficient to
counter the negative impact of both the credit crunch, and the macroeconomic
impact of debt reduction on aggregate demand.}

\FRAME{ftbphF}{4.0413in}{3.0536in}{0pt}{}{\Qlb{CreditCrunchAssets}}{%
11assetscrunchmodel.jpg}{\special{language "Scientific Word";type
"GRAPHIC";maintain-aspect-ratio TRUE;display "USEDEF";valid_file "F";width
4.0413in;height 3.0536in;depth 0pt;original-width 4.6561in;original-height
3.5103in;cropleft "0";croptop "1";cropright "1";cropbottom "0";filename
'11AssetsCrunchModel.wmf';file-properties "XNPEU";}}

{Romania, having only just completed the transition from a socialist to a
market economy, may therefore have to prepare itself for yet another
economic shock as the global capitalist system becomes mired in a debt trap.
We can have some modern guidance as to what this might mean for the economy
from the experience of Japan, which as noted earlier, is an important
exception to the general debt to GDP rule indicated in Figure Two.}

{This is because though, from the graph, Japan has the lowest rate of growth
of debt relative to GDP, this is only because Japan entered a
debt{}-deflation seventeen years ago, when its Bubble Economy collapsed at
the end of the 1980s. Japan}'{s private debt to GDP ratio has fallen
substantially since 1990{---}though the OECD Stat figures are not directly
comparable to the Australian and US data used above, they still imply a more
than 20 percent fall in the ratio since its peak. But this has been bought
at the cost of a seriously depressed economy, as shown by the OECD}'{s
Composite Leading Indicators (CLI), which indicate current living standards
and allow cross{}-country comparisons to be made.}

\FRAME{ftbphFU}{4.0413in}{3.2949in}{0pt}{\Qcb{Data Sources: \protect\cite%
{OECDData}}}{\Qlb{JapanDebtCLI}}{12japandebtdeflation.jpg}{\special{language
"Scientific Word";type "GRAPHIC";maintain-aspect-ratio TRUE;display
"USEDEF";valid_file "F";width 4.0413in;height 3.2949in;depth
0pt;original-width 4.6458in;original-height 3.781in;cropleft "0";croptop
"1";cropright "1";cropbottom "0";filename
'12JapanDebtDeflation.wmf';file-properties "XNPEU";}}

{Whereas Japan was far ahead of the USA on the CLI scale in 1990, when the
Bubble Economy collapsed, Japan}'{s economic performance stalled{---}and by
2004, American living standards had moved ahead of Japan}'{s. Japanese
living standards today are no better than they were a decade earlier{---}%
whereas previously, living standards improved by up to 50 percent every
decade. There are doubtless other factors that have contributed to this
stagnation, but the debt crisis of the early 1990s and its long{}-running
aftermath are key to understanding why, in 1990, the sun stopped rising in
the land of the rising sun.}

{When Japan}'{s crisis occurred, many Western economists blamed it on the
lack of transparency of the Western (and in particular, American) financial
system, and argued that Japan}'{s crisis simply couldn}'{t occur in the
West, because of its higher financial standards and superior financial
innovation. Today}'{s crisis may show that this was simply American hubris:
Japan}'{s 1990s crisis was caused by a speculative bubble focused on real
estate, and America has clearly followed suit in the subsequent decades.}

\section{No way to run an economy}

{Though a market economy is clearly preferable to a centrally planned one
from Romania}'{s own experience, there is something somewhat insane in a
system that allows itself to be periodically despoiled by pointless
speculation on housing, and excessive debt accumulation. Is this phenomenon
something that one must accept as inevitably a part of a capitalist system,
or is there something that could be done to stop, or at least attenuate,
this process in the future?}

{There are ways in which some financial instability in capitalism is
inevitable. As Minsky once remarked, {"}stability is destabilizing{"},
because a period of tranquil growth leads investors to revise their risk
expectations, thus leading to accelerating growth and the accumulation of
debt.}

{However, much of the long{}-term runup in debt has nothing to do with
actual investment, but instead involves pure speculation on asset prices.
This in turn is possible because, when an asset bubble takes hold, a
disconnect develops between asset prices and the income flows those assets
generate: share prices rise much faster than corporate earnings, while house
prices rise faster than the rents. We need some means to attenuate the
willingness of banks to fund speculation on shares and houses when such
bubbles arise.}

{One possible means with housing would be to limit the security that a
lender can get over a house to the income that the house itself can be
expected to generate.}

{At present, if a borrower defaults on a mortgage, then the lender gets
ownership of the house, and can sell it for whatever it can fetch on the
market. However, it would be possible to reform lending laws so that the
financial limit of the security the lender received was some sensible
multiple of the (actual or imputed) rental income from the property. Lenders
would still be able to lend as much as they wished to a borrower, but would
share some of the risk if he overextended himself. It is feasible that this
sharing of risk could reduce the {"}moral hazard{"} that the current system
generates, and thereby reduce the supply of credit to fund real estate
speculation.}

{While this is not as radical a change as the move from a socialist to a
capitalist economy that Romania has already undertaken, such changes tend to
be very hard to achieve in a capitalist economy because of the political
power of lenders, and the many vested interests in real estate. However
surely such a change is worth contemplating if the alternative is a fiasco
like the USA}'{s current subprime crisis, and an extended Depression like
that which still dominates Japan today.}

\end{document}